\documentclass[aoas,MSNbibl,nameyear,dvips]{arximspdf}
\usepackage{dcolumn}
\usepackage{graphicx}

\doi{10.1214/12-AOAS578} 
\volume{7}
\issue{1}
\pubyear{2013}
\firstpage{269}
\lastpage{294}

\makeatletter

\newcolumntype{d}[1]{D{.}{.}{#1}}
\newcommand{\tr}{\operatorname{tr}}

\newcommand{\VV}[1]{\mathbf{#1}}
\newcommand{\M}[1]{\bolds{#1}}
\newcommand{\MM}[1]{\mathbf{#1}}
\newcommand{\diag}{\operatorname{diag}}
\setattribute{abstract}   {width}  {290pt}
\makeatother

\begin{document}
\begin{frontmatter}

\title{Sparse integrative clustering of multiple omics~data sets}
\runtitle{Sparse iCluster}

\begin{aug}
\author[A]{\fnms{Ronglai} \snm{Shen}\corref{}\thanksref{t3}\ead[label=e1]{shenr@mskcc.org}},
\author[B]{\fnms{Sijian} \snm{Wang}\ead[label=e2]{swang@biostat.wisc.edu}}
\and
\author[C]{\fnms{Qianxing} \snm{Mo}\ead[label=e3]{qmo@bcm.edu}}
\thankstext{t3}{Supported by a Starr Cancer Consortium grant and
NIH Grant U24 CA143840.}
\runauthor{R. Shen, S. Wang and Q. Mo}
\affiliation{Memorial
Sloan-Kettering Cancer Center,
University~of~Wisconsin-Madison and~Baylor~College of Medicine}
\address[A]{R. Shen\\
Department of Epidemiology\\
\quad and Biostatistics\\
Memorial Sloan-Kettering Cancer Center\\
New York, New York 10065\\
USA\\
\printead{e1}}

\address[B]{S. Wang\\
Department of Biostatistics\\
\quad and Medical Informatics\\
Department of Statistics\\
University of Wisconsin-Madison\\
Madison, Wisconsin 53792-4675\\
USA\\
\printead{e2}}

\address[C]{Q. Mo\\
Division of Biostatistics\\
Dan L. Duncan Cancer Center\\
Baylor College of Medicine\\
Houston, Texas, 77030\\
USA\\
\printead{e3}}
\end{aug}

\received{\smonth{2} \syear{2012}}
\revised{\smonth{6} \syear{2012}}

%
\begin{abstract}
High resolution microarrays and second-generation sequencing platforms
are powerful tools to investigate genome-wide alterations in DNA copy
number, methylation and gene expression associated with a disease. An
integrated genomic profiling approach measures multiple omics data
types simultaneously in the same set of biological samples. Such approach renders
an integrated data resolution that would not be available with
any single data type. In this study, we use penalized latent variable
regression methods for joint modeling of multiple omics data types to
identify common latent variables that can be used to cluster patient
samples into biologically and clinically relevant disease subtypes. We
consider lasso [\textit{J. Roy. Statist. Soc. Ser. B} \textbf{58}
(1996) 267--288], elastic net
[\textit{J. R. Stat. Soc. Ser. B Stat. Methodol.} \textbf{67} (2005)
301--320] and
fused lasso [\textit{J. R. Stat. Soc. Ser. B Stat. Methodol.} \textbf
{67} (2005) 91--108]
methods to induce sparsity in the
coefficient vectors, revealing important genomic features that have
significant contributions to the latent variables. An iterative ridge
regression is used to compute the sparse coefficient vectors. In model
selection, a uniform design
[\textit{Monographs on Statistics and Applied Probability} (1994)
Chapman \& Hall] is used to seek
``experimental'' points that scattered uniformly across the search
domain for efficient sampling of tuning parameter combinations. We
compared our method to sparse singular value decomposition (SVD) and
penalized Gaussian mixture model (GMM) using both real and simulated
data sets. The proposed method is applied to integrate genomic,
epigenomic and transcriptomic data for subtype analysis in breast and
lung cancer data sets.\vspace*{-1pt}
\end{abstract}

%

\end{frontmatter}

\section{Introduction}\label{sec1}
Clustering analysis is an unsupervised learning method that aims to
group data into distinct clusters based on a certain measure of
similarity among the data points.\vadjust{\goodbreak} Clustering analysis has many
applications in a wide variety of fields including pattern recognition,
image processing and bioinformatics. In gene expression microarray
studies, clustering cancer samples based on their gene expression
profile has revealed molecular subgroups associated with
histopathological categories, drug response and patient survival
differences [\citet{Perou1999, Alizadeh2000, Sorlie2001, Lapointe2003,
Hoshida2009}].

In the past few years, \textit{integrative genomic studies} are emerging
at a fast pace where in addition to gene expression data, genome-wide
data sets capturing somatic mutation patterns, DNA copy number
alterations and DNA methylation changes are simultaneously obtained in
the same biological samples. A fundamental challenge in translating
cancer genomic findings into clinical application lies in the ability
to find ``driver'' genetic and genomic alterations that contribute to
tumor initiation, progression and metastasis [\citet{ChinGray2008,
simon2010}]. As integrated genomic studies have emerged, it has become
increasingly clear that true oncogenic mechanisms are more visible when
combining evidence across patterns of alterations in DNA copy number,
methylation, gene expression and mutational profiles [\citet{TCGA2008,
TCGAOV}]. Integrative analysis of multiple ``omic'' data types can help
the search for potential ``drivers'' by uncovering genomic features that
tend to be dysregulated by multiple mechanisms [\citet{ChinGray2008}]. A
well-known example is the \textit{HER2} oncogene which can be activated
through DNA amplification and mRNA over-expression. We will discuss the
\textit{HER2} example further in our motivating example.

In this paper, we focus on the class discovery problem given multiple
omics data sets (multidimensional data) for tumor subtype discovery.
A~major challenge in subtype discovery based on gene expression
microarray data is that the clinical and therapeutic implications for
most existing molecular subtypes of cancer are largely unknown. A
confounding factor is that expression changes may be related to
cellular activities independent of tumorigenesis, and therefore leading
to subtypes that may not be directly relevant for diagnostic and
prognostic purposes. By contrast, as we have shown in our previous work
[\citet{Shen5}], a joint analysis of multiple omics data
types offers a new paradigm to gain additional insights. Individually,
none of the genomic-wide data type alone can completely capture the
complexity of the cancer genome or fully explain the underlying disease
mechanism. Collectively, however, true oncogenic mechanisms may emerge
as a result of joint analysis of multiple genomic data types.

Somatic DNA copy number alterations are key characteristics of cancer
[\citet{Beroukhim2010}]. Copy number gain or amplification may lead to
activation of oncogenes (e.g., \textit{HER2} in Figure~\ref{fig1}). Tumor
suppressor genes can be inactivated by copy number loss.
High-resolution array-based comparative genomic hybridization (aCGH)
and SNP arrays have become dominant platforms for generating\vadjust{\goodbreak}
genome-wide copy number profiles. The measurement typical of aCGH
platforms is a log-ratio of normalized intensities of genomic DNA in
experimental versus control samples. For SNP arrays, copy number
measures are represented by log of total copy number (logR) and
parent-specific copy number as captured by a B-allele frequency (BAF)
[\citet{Chen2011,Olshen2011}]. Both platforms generate contiguous copy
number measures along ordered chromosomal locations (an example is
given in Figure~\ref{fig6}). Spatial smoothing methods are desirable for
modeling copy number data.

In addition to copy number aberrations, there are widespread DNA
methylation changes at CpG dinucleotide sites (regions of DNA where a
Cytocine nucleotide occurs next to a Guanine nucleotide) in the cancer
genome. DNA methylation is the most studied epigenetic event in cancer
[\citet{Holliday1979,Feinberg1983}, Laird (\citeyear{Laird2003,Laird2010})]. Tumor
suppressor genes are frequently inactivated by hypermethylation
(increased methylation of CpG sites in the promoter region of the
gene), and oncogenes can be activated through the promoter
hypomethylation. DNA methylation arrays measure the intensities of
methylated probes relative to unmethylated probes for tens of thousands
of CpG sites located at promoter regions of protein coding genes.
M-values are calculated by taking log-ratios of methylated and
unmethylated probe intensities [\citet{CHARM}], similar to the M-values
used for gene expression microarrays which quantify the relative
expression level (abundance of a gene's mRNA transcript) in cancer
samples compared to a normal control.

In this paper, we focus on the class discovery problem given multiple
omics data sets for tumor subtype discovery.
Suppose $t=1,\ldots,T$ different genome-scale data types (DNA copy
number, methylation, mRNA expression, etc.) are
obtained in $j=1,\ldots,n$ tumor samples. Let $\MM X_t$ be the $p_t
\times n$ data matrix where $\VV x_{it}$ denote
the $i$th row and $\VV x_{jt}$ the $j$th column of $\MM X_t$. Rows are
genomic features and columns are samples.
Here we use the term \textit{genomic feature} and the corresponding
feature index $i$ in the equations throughout
the paper to refer to either a protein-coding gene (typically for
expression and methylation data) or ordered
genomic elements that do not necessarily have a one-to-one mapping to a
specific gene (copy number measure
along chromosomal positions) depending on the data type.

Let $\MM Z$ be a $g \times n$ matrix where rows are latent variables and
columns are samples, and $g$ is the number of latent variables. Latent
variables can be interpreted as ``fundamental'' variables that determine
the values of the original $p$ variables [\citet{Jolliffe2002}]. In our
context, we use latent variables to represent disease driving factors
(underlying the wide spectrum of genomic alterations of various types)
that determine biologically and clinically relevant subtypes of the
disease. Typically, $g \ll\sum_t p_t$, providing a low-dimension
latent subspace to the original genomic feature space. Following a
similar argument for reduced-rank linear discriminant analysis in \citet
{Elements}, a rank-$g$\vadjust{\goodbreak} approximation where $g \le K-1$ is sufficient
for separating $K$ clusters among the $n$ data points. For the rest of
the paper, we assume the dimension of $\MM Z$ is $(K-1) \times n$ with
mean zero and identity covariance matrix. A joint latent variable model
expressed in matrix form is
%
%
\begin{equation}
\label{eqmodel} \MM X_t= \MM W_t\MM Z+\MM E_t,\qquad
t=1,\ldots, T.
\end{equation}
In the above, $\MM W_t$ is a $p_t \times(K-1)$ coefficient (or loading)
matrix relating $\MM X_t$ and $\MM Z$ with $\VV w_{jt}$ being the $j$th
row and $\VV w_{kt}$ the $k$th column of $\MM W_t$, and $\MM E_t$ is a
$p_t \times n$ matrix where the column vectors $\VV e_j, j=1,\ldots,n$,
represent uncorrelated error terms that follow a multivariate
distribution with mean zero and a diagonal covariance matrix $\M\Psi_t=(\sigma^2_{1},\ldots,\sigma^2_{p_t})$. Each data matrix is
row-centered so no intercept term is presented in equation (\ref
{eqmodel}).

Equation (\ref{eqmodel}) provides an effective integration framework in which the
latent variables $\MM Z=(\VV z_1, \ldots, \VV z_{K-1})$ are common for all
data types, representing a probabilistic low-rank approximation
simultaneously to the $T$ original data matrices. In Section~\ref{sec3.2} we
point out its connection and differences from singular value
decomposition (SVD). In Sections~\ref{sec6} and~\ref{sec7} we illustrate that applying
SVD to the combined data matrix broadly fails to achieve an effective
integration of various data types.

Equation (\ref{eqmodel}) is the basis of our initial work [\citet{Shen5}] in which
we introduced an integrative model called iCluster. We considered a
soft-thresholding estimate of $\MM W_t$ that continuously shrinks the
coefficients for noninformative features toward zero. The motivation
for sparse coefficient vectors is clearly indicated by Figure~\ref{fig1} panels
(D) and (E). A basic sparsity-inducing approach is to use a lasso penalty
[\citet{lasso1996}]. Nevertheless, different data types call for
appropriate penalty terms such that each $\MM W_t$ is sparse with a
specific sparsity structure. In particular, copy number aberrations
tend to occur in contiguous regions along chromosomal positions (Figure
\ref{fig6}), for which the fused lasso penalty [\citet{fusedlasso2005}] is
appropriate. In gene expression data where groups of genes involved in
the same biological pathway are co-regulated and thus highly correlated
in their expression levels, the elastic net penalty [\citet{enet2005}]
is useful to encourage a grouping effect by selecting strongly
correlated features together. In this paper, we present a sparse
iCluster framework that employs different penalty terms for the
estimation of $\MM W_t$ associated with different data types.

In Section~\ref{sec3} we present the methodological details of the latent
variable regression combined with lasso, elastic net and fused lasso
penalty terms. To determine the optimal combination of the penalty
parameter values, a very large search space needs to be covered, which
presents a computational challenge. An exhaustive grid search is
ineffective. We use a uniform design by Fang and Wang (\citeyear{glp1994}) that seeks
``experimental'' points that scattered uniformly across the search
domain has superior convergence rates over the conventional grid search\vadjust{\goodbreak}
(Section~\ref{sec3.3}). Section~\ref{sec4} presents an EM algorithm for maximizing the
penalized data log-likelihood. The number of clusters $K$ is unknown
and must be estimated. Section~\ref{sec5} discusses the estimation of $K$ based
on a cross-validation approach. Section~\ref{sec6} presents results from
simulation studies. Section~\ref{sec7} presents results from real data
applications. In particular, Section~\ref{sec7.1} presents an integrative
analysis of epigenomic and transcriptomic profiling data using a breast
cancer data set [\citet{Holm}]. In Section~\ref{sec7.2} we illustrate our
proposed method to construct a genome-wide portrait of copy number
induced gene expression changes using a lung cancer data set [\citet
{Ladanyi}]. We conclude the paper with a brief summary in Section~\ref{sec8}.

\section{Motivating example}\label{sec2}
In this section we show an example where an integrated analysis of
multiple omics data sets is far more insightful than separate analyses.
\citet{Pollack2002} used customized microarrays to generate
measurements of DNA copy number and mRNA expression in parallel for 37
primary breast cancer and 4 breast cancer cell line samples. Here the
number of data types $T=2$. In the mRNA expression data matrix $\MM
X_1$, the individual element $x_{ij1}$ refers to the observed
expression of the $i$th gene in the $j$th tumor. In the DNA copy number
data matrix $\MM X_2$, the individual element $x_{ij2}$ refers to the
observed log-ratio of tumor versus normal copy number of the $i$th gene
in the $j$th tumor. In this example, both data types have gene-centric
measurement by design.

\begin{figure}

\includegraphics{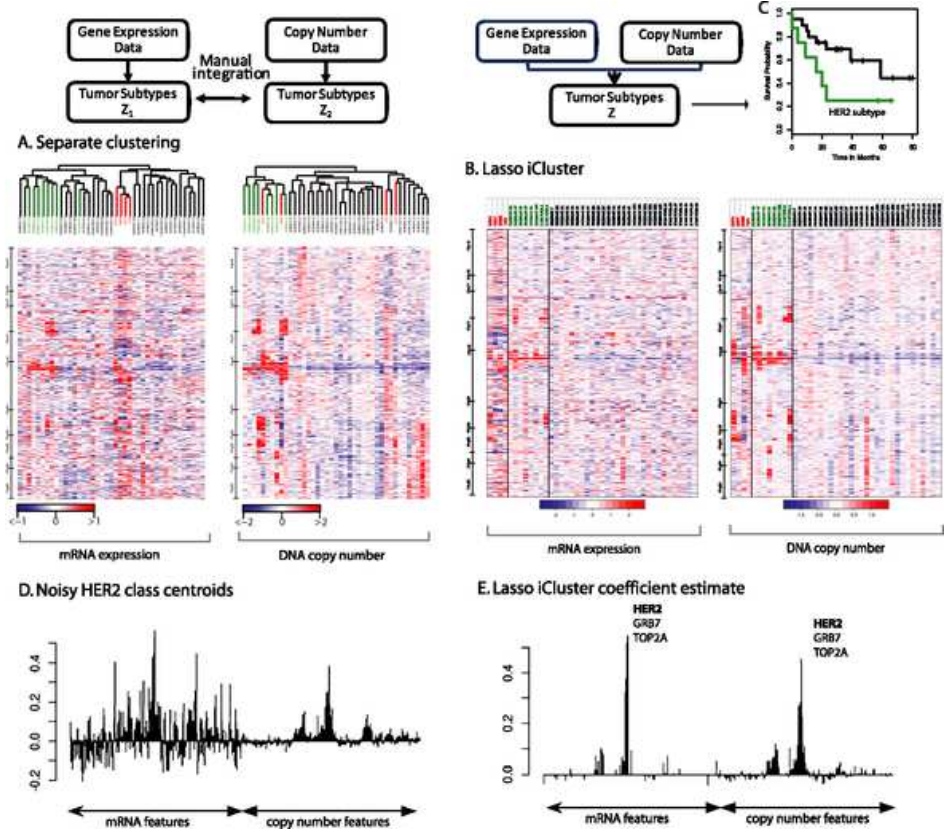}\vspace*{-3pt}

\caption{A motivating example using the Pollack data set to demonstrate
that a joint analysis using the lasso iCluster method outperforms the
separate clustering approach in subtype analysis given DNA copy number
and mRNA expression data. \textup{(A)} Heatmap with samples ordered by separate
hierarchical clustering. Rows are genes and samples are columns.
Samples labeled in red are breast cancer cell line samples. Samples
labeled in green are HER2 breast tumors. \textup{(B)} Heatmap with samples
ordered by integrative clustering using the lasso iCluster method. \textup{(C)}
Kaplan--Meier plot indicates the HER2 subtype has poor survival outcome.
\textup{(D)} Standard cluster centroid estimates. \textup{(E)} Sparse coefficient
estimates under the lasso iCluster model.}\label{fig1}\vspace*{-3pt}
\end{figure}

A heatmap of the genomic features on chromosome 17 is plotted in Figure~\ref{fig1}.
In the heatmap, rows are genes ordered by their genomic position and
columns are samples ordered by hierarchical clustering [panels (A)] or by
the lasso iCluster method [panels (B)]. There are two main subclasses in
the 41 samples: the cell line subclass (samples labeled in red) and the
\textit{HER2} tumor subclass (samples labeled in green). It is clear in Figure
\ref{fig1}(A) that these subclasses cannot be distinguished well from separate
hierarchical clustering analyses.

%

Separate clustering followed by manual integration as depicted in
Figure~\ref{fig1}(A) remains the most frequently applied approach to analyze
multiple omics data sets in the current literature due to its
simplicity and the lack of a truly integrative approach. However,
Figure~\ref{fig1}(A) clearly shows its lack of a unified system for cluster
assignment and poor correlation of the outcome with biological and
clinical annotation. As we will illustrate in the simulation study in
Section~\ref{sec7}, separate clustering can fail drastically in estimating the
true number of clusters, classifying samples to the correct clusters
and selecting cluster-associated features. Several limitations of this
common approach are responsible for its poor performance:
\begin{itemize}
\item Correlation between data sets is not utilized to inform the
clustering analysis, ignoring an important piece of information that
plays a key role for identifying ``driver'' features of biological
importance.\vadjust{\goodbreak}
\item Separate analysis of \textit{paired} genomic data sets is an
inefficient use of the available information.
\item It is not straightforward to integrate the multiple sets of
cluster assignments that are data-type dependent without extensive
prior biological information.
\item The standard clustering method includes all genomic features
regardless of their relevance to clustering.
\end{itemize}

Our method aims to overcome these obstacles by formulating a joint
analysis across multiple omics data\vadjust{\goodbreak} sets. The heatmap in Figure
\ref{fig1}(B)
demonstrates the superiority of our working model in correctly
identifying the subgroups (vertically divided by solid black lines).
From left to right, cluster 1 (samples labeled in red) corresponds to
the breast cancer cell line subgroup, distinguishing cell line samples
from tumor samples. Cluster 2 corresponds to the \textit{HER2} tumor
subtype (samples labeled in green), showing concordant amplification in
the DNA and overexpression in mRNA at the \textit{HER2} locus (chr 17q12).
This subtype is associated with poor survival as shown in Figure~\ref{fig1}(C).
Cluster 3 (samples labeled in black) did not show any distinct
patterns, though a pattern may have emerged if there were additional
data types such as DNA methylation.

The motivation for sparseness in the coefficient estimates is
illustrated by Figure~\ref{fig1}(E). It clearly reveals the \textit{HER2}-subtype
specific genes (including \textit{HER2\textup{,} GRB7\textup{,} TOP2A}). By contrast, the
standard cluster centroid estimation is flooded with noise [Figure~\ref{fig1}(D)],
revealing an inherent problem with clustering methods without
regularization.

The copy number data example in Figure~\ref{fig1} depicts a narrow (focal) DNA
amplification event on a single chromosome involving only a few genes
(including \textit{HER2}). Nevertheless, copy number is more frequently
altered across long contiguous regions. In the lung cancer data example
we will discuss in Section~\ref{sec7.2}, chromosome arm-level copy number gains
(log-ratio~$>0$) and losses (log-ratio~$<0$) as illustrated in Figure~\ref{fig6}
are frequently observed, motivating the use of a fused lasso penalty to
account for such structural dependencies. In the next section we
discuss methodological details on lasso, fused lasso and elastic net in
the latent variable regression.

\section{Method}\label{sec3}
Assuming Gaussian error terms, equation (\ref{eqmodel}) implies the following
conditional distribution
%
%
\begin{equation}\label{eq2}
\MM X_t|\MM Z \sim N(\MM W_t \MM Z, \M
\Psi_t),\qquad  t=1,\ldots,T.
\end{equation}
Further assuming $\MM Z \sim N(\VV0,\MM I)$, the marginal distribution
for the observed data is then
%
%
\begin{equation}\label{eq3}
\MM X_t \sim N(\MM0, \M\Sigma_t),
\end{equation}
where $\bolds{\Sigma}_t=\MM W_t \MM W_t'+\M\Psi_t$. Direct maximization of the
marginal data log-likelihood is difficult. We consider an
expectation--maximization (EM) algorithm [\citet{Dempster1977}]. In the
EM framework, the latent variables are considered ``missing data.''
Therefore, the ``complete'' data log-likelihood that consists of these
latent variables is
%
%
\begin{eqnarray}\label{eq4}
\label{eqlc} \ell_c &=& -\frac{n}{2}\sum
_{t=1}^T\log|\M{\Psi}_t|-
\frac{1}{2}\sum_{t=1}^T \tr \bigl((\MM
X_t-\MM W_t\MM Z)'\M{\Psi}_t^{-1}(
\MM X_t-\MM W_t\MM Z)\bigr)
\nonumber
\\[-8pt]
\\[-8pt]
\nonumber
&&{}-\frac{1}{2}\tr \bigl(\MM
Z'\MM Z\bigr).
\end{eqnarray}
The constant term in $ \ell_c$ has been omitted. In the next section we
discuss a penalized complete data log-likelihood to induce sparsity in
$\MM W_t$.

\subsection{Penalized likelihood approach}\label{sec3.1}
As mentioned earlier, sparsity in $\MM W_t$ directly impacts the
interpretability of the latent variables. A zero entry in the $i$th row
and $k$th column ($w_{ikt}=0$) means that the $i$th genomic feature has
no weight on the $k$th latent variable in data type $t$. If the entire
row $\VV w_{it}=0$, then this genomic feature has no contribution to the
latent variables and is considered noninformative. We use a penalized
complete-data log-likelihood as follows to enforce desired sparsity in
the estimated
$\MM W_t$:
%
%
\begin{equation}\label{eq5}
\label{eqlcp} \ell_{c,p}\bigl(\{\MM{W}_t
\}_{t=1}^T,\{\bolds{\Psi}\}_{t=1}^T
\bigr)=\ell_{c}- \sum_{t=1}^T
J_{\lambda_t}(\MM{W}_t),
\end{equation}
where $\ell_c$ is the complete-data log-likelihood function defined in
(\ref{eqlc}) which controls the fitness of the model; $J_{\lambda_t}(\MM
{W}_t)$ is a penalty function which controls the complexity of the
model; and
$\lambda_t$ is a nonnegative tuning parameter that determines
the balance between the two. The subscript $p$ in $\ell_{c,p}$ stands
for penalized.

Different data types call for different penalty functions. We introduce
three types of penalties in the iCluster model: lasso, elastic net, and
fused lasso. Both lasso and elastic net regression methods have been
applied to gene expression data [\citet{Simon, ccle}]. For feature
selection, the elastic net may have an additional advantage by
shrinking coefficients of correlated features toward each other, and
thus encourages a grouping effect toward selecting highly correlated
features together. Copy number aberrations tend to occur in contiguous
regions along chromosomal positions, motivating the use of fused lasso.

\subsubsection{The lasso penalty}\label{sec3.1.1}
The lasso penalty is a basic sparsity-inducing that takes the form
%
%
\begin{equation}\label{eq6}
J_{\lambda_t}(\MM{W}_t)=\lambda_t\sum
_{k=1}^{K-1}\sum_{i=1}^{p_t}|w_{ikt}|,
\end{equation}
where $w_{ikt}$ is the element in the $i$th row and $k$th column of $\MM
W_t$. The $\ell_1$-penalty continuously shrinks the coefficients toward
zero and thereby yields a substantial decrease in the
variance of the coefficient estimates. Owing to the singularity of the
$\ell_1$-penalty at the origin ($w_{ikt}=0$), some estimated
$\hat{w}_{ikt}$ will be \textit{exactly} zero. The degree of sparseness
is controlled by the tuning parameter $\lambda_t$.

\subsubsection{The fused lasso penalty}\label{sec3.1.2}
To account for the strong spatial dependence along genomic ordering
typical in DNA copy number data, we consider the fused lasso
penalty\vadjust{\goodbreak}
[\citet{fusedlasso2005}],
which takes the following form:
%
%
\begin{equation}\label{eq7}
\label{eqfusedlassi} J_{\lambda_t}(\MM{W}_t)=\lambda_{1t}
\sum_{k=1}^{K-1}\sum
_{i=1}^{p_t}|w_{ikt}|+\lambda_{2t}
\sum_{k=1}^{K-1}\sum
_{i=2}^{p_t}|w_{ikt}-w_{(i-1)kt}|,
\end{equation}
where $\lambda_{1t}$ and $\lambda_{2t}$ are two nonnegative tuning
parameters. The first penalty encourages sparseness while the second
encourages smoothness along index~$i$. The fused lasso penalty is
particularly suitable for DNA copy number data where contiguous regions
of a chromosome tend to be altered in the same fashion [\citet
{TibshiraniWang2008}].

\subsubsection{The elastic net penalty}\label{sec3.1.3}
The elastic net penalty [Zou and Hastie (\citeyear{enet2005})] takes the form
%
%
\begin{equation}\label{eq8}
\label{eqenet} J_{\lambda_t}(\MM{W}_t)=\lambda_{1t}\sum
_{k=1}^{K-1}\sum_{i=1}^{p_t}|w_{ikt}|+
\lambda_{2t}\sum_{k=1}^{K-1}\sum
_{i=1}^{p_t}w_{ikt}^2,
\end{equation}
where $\lambda_{1t}$ and $\lambda_{2t}$ are two nonnegative tuning
parameters. Zou and Hastie (\citeyear{enet2005}) showed that the elastic net
penalty tends to select or remove highly correlated predictors
together in a linear regression setting by enforcing their estimated
coefficients to be similar. In our experience, the elastic
net penalty tends to be more numerically stable than the lasso penalty
in our model.

\begin{figure}

\includegraphics{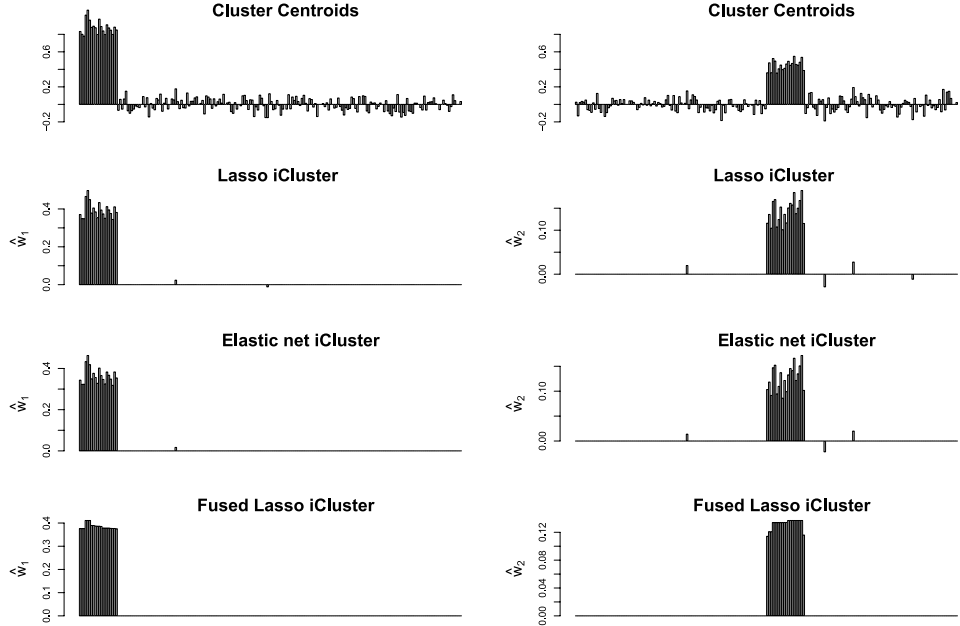}

\caption{A simulated pair of data sets each with 100 subjects ($n=100$)
and 200 features ($p_t = 200, t=1,2$), and 2 subgroups ($K=2$). Top
panel plots the cluster centroids in data set 1 (left) and in data set
2 (right). Estimated sparse iCluster coefficients are plotted
below.}\label{fig2}
\end{figure}

Figure~\ref{fig2} shows the effectiveness of sparse iCluster using a simulated
pair of data sets ($T=2$). We simulated a single length-n latent
variable $\VV z \sim N(\VV0,\MM I)$ where $n=100$. The coefficient matrix
$\MM W_1$ consists of a single column $\VV w_1$ of length $p_1=200$ with
the first 20 elements set to 1.5 and the remaining elements set to~0,
that is, $w_{i1}=1.5$ for $i=1,\ldots, 20$ and 0 elsewhere. The
coefficient matrix $\MM W_2$ consists of a single column $\VV w_2$ of
length $p_2=200$ and set to have $w_{i2}=1.5$ for $i=101,\ldots, 120$
and 0 elsewhere. The lasso, elastic net and fused lasso coefficient
estimates are plotted to contrast the noisy cluster centroids estimated
separately in data type 1 (left) and in data type~2 (right) in the top
panel of Figure~\ref{fig2}. The algorithm for computing these sparse estimates
will be discussed in Section~\ref{sec4}.

%

\subsection{Relationship to singular value decomposition (SVD)}\label{sec3.2}
An SVD/PCA on the concatenated data matrix $\MM X=(\MM X_1', \ldots, \MM
X_T')'$ is a special case of equation (\ref{eqmodel}) that requires a common
covariance matrix across data types. Specifically, it can be shown that
when $\M\Psi_1=\cdots=\M\Psi_T=\sigma^2 \MM I$, equation (\ref{eqmodel}) reduces
to a ``probabilistic SVD/PCA'' on the concatenated data matrix $\MM X$.
Following similar derivation in Tipping and Bishop (\citeyear{TB99}), the maximum
likelihood estimates of $\MM W$, where $\MM W=(\MM W_1',\ldots,\MM W_T')'$
is the concatenated coefficient matrix, coincide with the first $K-1$
eigenvectors of the sample covariance matrix $\MM X\MM X'$ or the right
singular vector of the concatenated data matrix~$\MM X$. The MLE of
$\sigma^2$ is the average of the remaining $n-K+1$ eigenvalues,
capturing the residual variation averaged over the ``lost''
dimensions.\vadjust{\goodbreak}

The major assumption is the requirement that all features have the same
variance. The genomic data types, however, are fundamentally different
and the method we propose primarily aims to deal with
heteroscedasticity among genomic features of various types. The common
covariance assumption that leads to SVD is therefore not suitable for
integrating omics data types. It is worth mentioning that feature
scaling may not necessarily yield $\sigma^2_1=\cdots=\sigma^2_{p_t}$.
In our modeling framework, $\sigma^2_i$ is the conditional variance of
$x_{ij}$ given $\VV z_j$. Standardization on $x_{ij}$ will yield the
same marginal variance across features, but the conditional variances
of features are not necessarily the same after standardization.

Our method aims to identify common influences across data types through
the latent component $\MM Z$. The independent error terms $\MM E_t,
t=1,\ldots,T$ capture the remaining variances unique to each data type
after accounting for the common variance. In SVD, however, the unique
variances are absorbed in the term $\MM W \MM Z$ by enforcing
$\M\Psi_1=\cdots=\M\Psi_T=\sigma^2 \MM I$. As a result, common and unique
variations are no longer separable. This is in fact one of the
fundamental differences between the factor analysis model and PCA,
which has practical importance in integrative modeling.

In Sections~\ref{sec6} and~\ref{sec7} we illustrate that SVD on the concatenated data
matrix broadly fails to achieve an effective integration in both
simulated and real data sets. By contrast, our method can more
effectively deal with heteroscedasticity among genomic features of
various types. The contrast with a sparse SVD method lies in the fact
that our framework allows block-wise sparse constraints to the
coefficient matrix.

\subsection{Uniform sampling}\label{sec3.3}
An exhaustive grid search for the optimal combination of the penalty
parameters that maximizes a certain criterion (the optimization
criterion will be discussed in Section~\ref{sec5}) is inefficient and
computationally prohibitive. We use the uniform design (UD) of \citet
{glp1994} to generate good lattice points from the search domain, a
similar strategy adopted by \citet{WangBJ2008}. A key theoretical
advantage of UD over the traditional grid search is the uniform space
filling property that avoids wasteful computation at close-by points.
Let $D$ be the search region. Using the concept of discrepancy that
measures uniformity on $D\subset R^d$ with arbitrary dimension $d$,
which is basically
the Kolmogorov statistic for a uniform distribution on $D$, Fang and Wang
(\citeyear{glp1994}) point out that the discrepancy of the good lattice point set
from a uniform design converges to zero with a rate of $O(n^{-1}(\log
n)^d)$, where $n$ (a prime number) denotes the number of generated
points on $D$. They also point out that the sequence of equi-lattice
points on $D$ has a rate of $O(n^{-1/d})$ and the sequence of uniformly
distributed random numbers on D has a rate of $O(n^{-1/2}(\log\log
n)^{1/2})$. Thus, the uniform design has an optimal rate for $d\ge2$.

\section{Algorithm}\label{sec4}
We now discuss the details of our algorithm for parameter estimation in
sparse iCluster. The latent variables (columns of $\MM Z$) are
considered to be ``missing'' data. The algorithm therefore iterates
between an E-step for imputing~$\MM Z$ and a penalized maximization step
(M-step) that updates the estimates of $\MM W_t$ and $\M\Psi_t$ for all
$t$. Given the latent variables, the data types are conditionally
independent and, thus, the integrative omics problem can be decomposed
into solving~$T$ independent subproblems with suitable penalty terms.
The penalized estimation procedures are therefore ``decoupled'' for each
data type given the latent variables~$\MM Z$. When convergence is
reached, cluster membership will be assigned for each tumor based on
the posterior mean of the latent variable $\MM Z$.

\begin{enumerate}
\item[\textit{E-step}.]
In the E-step, we take the expectation of the penalized complete-data
log-likelihood $\ell_{c,p}$ as defined in equations (\ref{eqlc}) and
(\ref{eqlcp}), which primarily involves computing two conditional
expectations given the current parameter estimates:
%
%
\begin{eqnarray}\label{eq9}
 E[\MM{Z}|\MM X]&=&\MM W'\M\Sigma^{-1}\MM X
\\
\label{eq10} E\bigl[\MM Z\MM Z'|\MM X\bigr]&=&\MM I-\MM W'\M
\Sigma^{-1}\MM W+E[\MM{Z}|\MM X]E[\MM {Z}|\MM X]',
\end{eqnarray}
where $\M\Sigma=\MM W \MM W'+\M\Psi$ and $\M\Psi=\diag(\M\Psi_1,
\ldots, \M\Psi_T)$. Here, the posterior mean in~(\ref{eq9}) effectively
provides a simultaneous rank-$(K-1)$ approximation to the original data
matrices $\MM X$.

\item[\textit{M-step}.]
In the M-step, given the quantities in equations (\ref{eq9}) and (\ref{eq10}), we
maximize the penalized complete-data log-likelihood to update the
estimates of $\MM W_t$ and~$\M\Psi_t$.

\begin{enumerate}[(1)]
\item[(1)] \textit{Sparse estimates of $\MM W_t$}.
For $t=1,\ldots,T$, we obtain the penalized estimates by
%
%
\begin{eqnarray}
\label{eqmstep} \MM W_t&\leftarrow&\mathop{\operatorname{argmin}}_{\MM{W}_t}
\frac
{1}{2}\sum_{t=1}^TE \bigl[\tr
\bigl((\MM{X}_{t}-\MM{W}_t\MM{Z})'\M{
\Psi}_t^{-1}(\MM {X}_{t}-\MM{W}_t
\MM{Z})\bigr) | \hat{\MM W}_t,\hat{\M\Psi}_t
\bigr]
\nonumber
\\[-8pt]
\\[-8pt]
\nonumber
&&{}+J_{\lambda
_t}(\MM{W}_t),
\end{eqnarray}
where $\hat{\MM W}_t$ and $\hat{\M\Psi}_t$ denote the parameter
estimates in the last EM iteration. We apply a local quadratic
approximation [\citet{FanLi2001}] to the $\ell_1$ term involved in the
penalty function $J_{\lambda_t}(\MM{W}_t)$. Using the fact $|\alpha
|=\alpha^2/|\alpha|$ when $\alpha\neq0$, we consider the following
quadratic approximation to the $\ell_1$ term:
%
%
\begin{equation}\label{eq12}
\lambda_t\sum_{k=1}^{K-1}\sum
_{i=1}^{p_t}\frac{w_{ikt}^2}{|\hat{w}_{ikt}|}.
\end{equation}

Due to the uncorrelated error terms (diagonal $\M\Psi_t$) and
``noncoupling'' structure of the lasso and elastic net penalty terms,
the estimation of $\MM W_t$ can then be computed feature-by-feature by
taking derivatives with respect to each row $\VV w_{it}$ for $i=1,\ldots
,p_t$. The solution for (\ref{eqmstep}) under various penalty terms
can then be obtained by iteratively computing the following ridge
regression estimates:

\begin{enumerate}[(a)]
\item[(a)] \textit{Lasso estimates}. For $i=1,\ldots,p_t$,
%
%
\begin{equation}
\label{eqQAlasso} \VV{w}_{it}=\VV x_{it} E \bigl[\MM
Z' |\MM X_t, \hat{\MM W}_t, \hat{\M\Psi
}_t \bigr] \bigl(E \bigl[\MM Z \MM Z' |\MM
X_t, \hat{\MM W}_t, \hat{\M\Psi }_t \bigr]
+\MM{A}_i \bigr)^{-1} ,
\end{equation}
where $\MM{A}_i=2\sigma_i^2\lambda_t\diag\{1/|\hat w_{i1t}|,\dots
,1/|\hat w_{i(K-1)t}|\}$. Computing (\ref{eqQAlasso}) only requires
the inversion of a $(K-1)\times(K-1)$ matrix in the latent subspace.

\item[(b)] \textit{Elastic net estimates}.
Similarly, we consider a quadratic approximation to the $\ell_1$ term
in the elastic net penalty
and obtain the solution for (\ref{eqmstep}) by iteratively computing a
ridge regression estimate similar to (\ref{eqQAlasso}) but with $\VV
{A}_i=2\sigma_i^2 (\lambda_{1t}\diag\{1/|\hat{w}_{i1t}|,\dots,1/\break|\hat
{w}_{i(K-1)t}|\}+\lambda_{2t}\VV{I} )$.

\item[(c)] \textit{Fused lasso estimates}.
For fused lasso penalty terms, we consider the following approximation:
%
%
\begin{equation}
\label{eqQAfl} \lambda_{1t}\sum_{k=1}^{K-1}
\sum_{i=1}^p\frac{w_{ikt}^2}{|\hat{w}_{ikt}|}+
\lambda_{2t}\sum_{k=1}^{K-1}\sum
_{i=2}^p\frac{(w_{ikt}-w_{(i-1)kt})^2}{|\hat
{w}_{ikt}-\hat{w}_{(i-1)kt}|}.
\end{equation}
In the fused lasso scenario, the parameters are coupled together and
the estimation of $\VV{w}_i$ are no longer separable. However, we
circumvent the problem by expressing the estimating equation in terms
of a vectorized form $\tilde{\VV w}_t=\mbox{vec}(\VV{W}'_t)=(\VV w_1,\ldots
,\VV w_{K-1})'$, a column vector of dimension $s=p_t\cdot(K-1)$ by
concatenating the columns of $\MM W_t$. Then (\ref{eqQAfl}) can be
expressed in the following form:
\[
\lambda_{1t}\tilde{\VV w}'_t\VV{A}\tilde{\VV
w}_t+\lambda_{2t}\tilde{\VV w}'_t
\VV{L}\tilde{\VV w}_t,
\]
where
\begin{eqnarray*}
\hspace*{-15pt}\VV{A}&=&\diag\bigl \{1/|\hat{w}_1|,\dots,1/|
\hat{w}_{s}| \bigr\},
\\
\hspace*{-15pt}\VV{L} &=& \VV{D}-\VV{M},
\\
\hspace*{-15pt}\VV{M} &=& \cases{
1/|\hat{w}_i-
\hat{w}_j|, &\quad $|i-j|=K-1,$
\vspace*{2pt}\cr
0, &\quad $\mbox{otherwise}$}\qquad
(s\times s\mbox{ dimension}),
\\
\hspace*{-15pt}\VV{D} &=& \diag\{d_1,\dots,d_s\} \mbox{ where } d_j
\mbox{ is the summation of the } j \mbox{th row of } \VV{M}.
\end{eqnarray*}

Letting
$\MM C=\MM X_t E [\MM Z' |\MM X_t, \hat{\MM W}_t, \hat{\M\Psi}_t
]$ and $ \MM{Q}=E [\MM Z\MM Z' |\MM X_t, \hat{\MM W}_t, \hat{\M\Psi
}_t ]$, the corresponding estimating equation is then
%
%
\begin{equation}
\label{eq15} \frac{\partial}{\partial\tilde{\VV w}}J(\tilde{\VV w})+\tilde{\VV {Q}}\tilde{\VV w}=\tilde{
\VV{C}},
\end{equation}
where
%
%
\begin{equation}
\label{eq16} \tilde{\VV{Q}}=\pmatrix{
\sigma_1^{-2}\VV{Q}& &
\vspace*{2pt}\cr
& \ddots&
\vspace*{2pt}\cr
& & \sigma_{p_t}^{-2}\VV{Q}},\qquad   \tilde{\VV{C}}=\pmatrix{
\sigma_1^{-2}\VV{c}'_1
\\\vspace*{2pt}\cr
\vdots
\vspace*{2pt}\cr
\sigma_{p_t}^{-2}\VV{c}'_{p_t}},
\end{equation}
where $\VV c_j$ is the $j$th row of $\MM C$. The solution for (\ref
{eqmstep}) under the fused lasso penalty is then computed by
iteratively computing
%
%
\begin{equation}\label{eq17}
\tilde{\VV w}_t= (\tilde{\VV{Q}}+2\lambda_{1t}\VV{A}+2
\lambda_{2t}\VV{L} )^{-1}\tilde{\VV{C}}.
\end{equation}
\end{enumerate}

\item[(2)] \textit{Estimates of $\M\Psi_t$}.
Finally, for $t=1,\ldots,T$, we update $\M\Psi_t$ in the M-step as follows:
%
%
\begin{equation}\label{eq18}
\M\Psi_t=\frac{1}{n} \diag\bigl( \MM X_t \MM
X'_t-\hat{\MM W}_t E \bigl[\MM Z|\{\MM
{X}_t\}_{t=1}^T, \{\hat{\MM W}_t
\}_{t=1}^T, \{\hat{\M\Psi}_t
\}_{t=1}^T \bigr]\MM X'_t \bigr).
\end{equation}
\end{enumerate}
\end{enumerate}

The algorithm iterates between the E-step and the M-step as described
above until convergence. Cluster membership will then be assigned by
applying a standard K-means clustering on the posterior mean $E[\MM Z|\MM
X]$. In other words, cluster partition in the final step is performed
in the integrated latent variable subspace of dimension $n \times
(K-1)$. Applying $k$-means on latent variables to obtain discrete cluster
assignment is commonly used in spectral clustering methods [\citet
{spectral1, spectral2}].

\section{Choice of tuning parameters}\label{sec5}
We use a resampling-based criterion for selecting the penalty
parameters and the number of clusters. The procedure entails repeatedly
partitioning the data set into a learning and a test set. In each
iteration, sparse iCluster (for a given $K$ and tuning parameter values)
will be applied to the learning set to obtain a classifier and
subsequently predict the cluster membership for the test set samples.
In particular, we first obtain parameter estimates from the learning
set. For new observations in the test data $\MM X^*$, we then compute
the posterior mean of the latent\vspace*{-1pt} variables $E[\MM Z|\MM X^*]=\hat{\MM
W}_{\ell}'\hat{\M\Sigma}_{\ell}^{-1}\MM X^*$, where $\hat{\MM W}_{\ell},
\hat{\M\Sigma}^{-1}_{\ell}$ denote parameter estimates from the
learning set. A K-means clustering is then applied to $E[\MM Z|\MM X^*]$
to partition the test set samples into $K$ clusters. Denote this as
partition~$C_1$. In parallel, the procedure applies an independent
sparse iCluster with the same penalty parameter values to the test set
to obtain a second partition $C_2$, giving the ``observed'' test sample
cluster labels. Under the true model, the predicted $C_1$ and the
``observed'' $C_2$ (regarded as the ``truth'') would have good agreement
by measures such as the adjusted Rand index. We therefore define a
reproducibility index (RI) as the median adjusted Rand index across all
repetitions. Values of RI close to 1 indicate perfect cluster
reproducibility and values of RI close to 0 indicate poor cluster
reproducibility. In this framework, the concepts of bias, variance and
prediction error that typically apply to classification analysis where
the true cluster labels are known now become relevant for clustering.
The idea is similar to the ``Clest'' method proposed by \citet
{Dudoit2002}, the prediction strength measure proposed by \citet
{Tibshirani2005} and the in-group proportion (IGP) proposed by
\citet{Kapp2007}.

\section{Simulation}\label{sec6}
In this section we present results from two simulation studies. In the
first simulation setup, we simulate a single length-$n$ latent variable
$\VV z \sim N(0,1)$ where $n=100$. Subject $j, j=1,\ldots,n$, belongs to
cluster 1 if $z_j>0$ and cluster 2 otherwise. For simplicity, the pair
of coefficient matrices ($\MM W_1, \MM W_2$) are of the same dimension
$200 \times1$ ($p_1=p_2=200$), with $w_{it}=3$ for $i=1,\ldots,20$ for
both data types ($t=1,2$) and zero elsewhere. Next we obtain the data
matrices ($\MM{X}_1,\MM{X}_2$) with each element generated according to
equation (\ref{eqmodel}) with standard normal error terms. This simulation
represents a scenario where an effective joint analysis of two data
sets should be expected to enhance the signal strength and thus improve
clustering performance.

%
%
\begin{table}
\caption{Clustering performance summarized over 50 simulated data sets
under\break setup 1 ($K=2$). Separate clustering methods have two sets of
numbers\break associated with model fit to each individual data type. Numbers
in parentheses\break are the standard deviations over 50 simulations}\label{tab1}
\begin{tabular*}{\textwidth}{@{\extracolsep{\fill}}lccc@{}}
\hline
&\multicolumn{1}{c}{\textbf{Percent of}}&&\\
&\multicolumn{1}{c}{\textbf{times
choosing}}&\multicolumn{1}{c}{\textbf{Cross-validation}}&\multicolumn{1}{c@{}}{\textbf{Cluster}}\\
\multicolumn{1}{@{}l}{\textbf{Method}} & \multicolumn{1}{c}{\textbf{the correct} $\bolds{K}$} &
\multicolumn{1}{c}{\textbf{error rate}} &
\multicolumn{1}{c@{}}{\textbf{reproducibility}} \\
\hline
{Separate K-means} & {58} &
{0.08 (0.04)} & {0.67 (0.17)}\\
{ } & {62} & {0.08 (0.04)} & {0.70 (0.19)} \\
{Concatenated K-means} & {50} &
{0.06 (0.04)} & {0.66 (0.19)} \\
{Separate sparse SVD} & {74} &
{0.07 (0.06)} & {0.71 (0.13)}\\
{ } & {76} & {0.07 (0.07)} & {0.72 (0.12)} \\
{Concatenated sparse SVD} & {78} &
{0.07 (0.08)} & {0.70 (0.12)} \\
{Separate AHP-GMM} & {38} &
{0.06 (0.04)} & {0.72 (0.15)}\\
{ } & {40} & {0.05 (0.04)} & {0.74 (0.14)} \\
{Concatenated AHP-GMM} & {46} &
{0.06 (0.04)} & {0.75 (0.13)} \\
{Lasso iCluster} & {90} &
{0.04 (0.02)} & {0.81 (0.08)} \\
{Enet iCluster} & {94} &
{0.03 (0.02)} & {0.85 (0.07)} \\
{Fused lasso iCluster} & {94} &
{0.03 (0.02)} & {0.83 (0.08)} \\
\hline
\end{tabular*}
\end{table}

\begin{table}[b]
\caption{Feature selection performance summarized over 50 simulated
data sets for $K=2$. There are a total of 20 true features simulated to
distinguish the two sample clusters}\label{tab2}
\begin{tabular*}{\textwidth}{@{\extracolsep{\fill}}lcccc@{}}
\hline
& \multicolumn{2}{c}{\textbf{Data 1}} & \multicolumn{2}{c@{}}{\textbf{Data 2}}
\\[-4pt]
& \multicolumn{2}{c}{\hrulefill} & \multicolumn{2}{c@{}}{\hrulefill} \\
\multicolumn{1}{c}{} & \textbf{True} & \textbf{False} & \textbf{True}
& \textbf{False} \\
\multicolumn{1}{@{}l}{\textbf{Method}} & \textbf{positives} & \textbf{
positives} & \textbf{positives} & \textbf{positives} \\
\hline
{Separate K-means} & -- & -- & -- & --\\
{Concatenated K-means} & -- & -- & -- & --\\
{Separate sparse SVD} & 18.7 (3.2) & 21.5 (37.7) &
18.8 (2.9) & 27.4 (43.6) \\
{Concatenated sparse SVD} & 14.0 (5.3) & 22.5 (16.1)
& 13.7 (5.2) & 22.8 (16.4) \\
{Separate AHP-GMM} & 19.6 (2.1) & 0.02 (0.16) & 19.1
(3.1) & 0 (0) \\
{Concatenated AHP-GMM} & 18.8 (3.6) & 0.02 (0.15) &
18.6 (4.0) & 0.02 (0.15) \\
{Lasso iCluster} & 20 (0) & 0.07 (0.3) & 20 (0) &
0.07 (0.3) \\
{Enet iCluster} & 20 (0) & 0.1 (0.3) & 20 (0) & 0.02
(0.1) \\
{Fused lasso iCluster} & 20 (0) & 0 (0) & 20 (0) & 0
(0) \\
\hline
\end{tabular*}
\end{table}

Table~\ref{tab1} summarizes the performances of each method in terms of the
ability to choose the correct number of clusters, cross-validated error
rates and cluster reproducibility. In Table~\ref{tab1} separate K-means methods
perform poorly in terms of the ability to choose the correct number of
clusters, cluster reproducibility and the cross-validation error rates
(with respect to the true simulated cluster membership). K-means on
concatenated data performs even worse, likely due to noise
accumulation. For sparse SVD, a cluster assignment step is needed. We
took a similar approach of applying K-means on the first $K-1$ right
singular vectors of the data matrix. Sparse SVD performs better than
simple K-means, though data concatenation does not seem to offer much
advantage. In this simulation scenario, AHP-GMM models show good
performance in feature selection (Table~\ref{tab2}), but appear to have a low
frequency of choosing the correct $K=2$. A common theme in this
simulation is that a data concatenation approach is generally
ineffective regardless of the clustering methods used. By contrast,
sparse iCluster methods achieved an effective integrative outcome
across all performance criteria.

Table~\ref{tab2} summarizes the associated feature selection performance. No
numbers are shown for the standard K-means methods, as they do not have
an inherent feature selection method. Among the methods, sparse
iCluster methods perform the best in identifying the true positive
features while keeping the number of false positives close to 0.

In the second simulation, we vary the setup as follows. We simulate 150
subjects belonging to three clusters ($K=3$). Subjects $j=1,\ldots,50$
belong to cluster 1, subjects $j=51,\ldots,100$ belong to cluster 2,
and subjects $j=101,\ldots,150$ belong to cluster 3. A total of $T=2$
data types ($\MM X_1, \MM X_2$) are simulated. Each has $p_1=p_2=500$
features. Here each data type alone only defines two clusters out of
the three. In data set 1, $x_{ij1}\sim N(2,1)$ for $i=1,\ldots,10$ and
$j=1,\ldots,50$, $x_{ij1}\sim N(1.5,1)$ for $i=491,\ldots,500$ and
$j=51,\ldots,100$, and $x_{ij1}\sim N(0,1)$ for the rest. In data set
2, $x_{ij2}=0.5*x_{ij1}+e$ where $e \sim N(0,1)$ for $i=1,\ldots,10$
and $j=1,\ldots,50$, $x_{ij2}\sim N(2,1)$ for $i=491,\ldots,500$ and
$j=101,\ldots,150$, and $x_{ij2}\sim N(0,1)$ for the rest. The first 10
features are correlated between the two data types. In Tables~\ref{tab3} and~\ref{tab4},
the sparse iCluster methods consistently outperform the other methods
in clustering and feature selection.

%
%
\begin{table}
\caption{Clustering performance summarized over 50 simulated data sets
under setup 2 ($K=3$)}\label{tab3}
\begin{tabular*}{\textwidth}{@{\extracolsep{\fill}}lccc@{}}
\hline
&\multicolumn{1}{c}{\textbf{Frequency of}}&&\\
&\multicolumn{1}{c}{\textbf{choosing the}}&\multicolumn{1}{c}{\textbf{Cross-validation}}&\multicolumn{1}{c@{}}{\textbf{Cluster}}\\
\multicolumn{1}{@{}l}{\textbf{Method}} & \multicolumn{1}{c}{\textbf{correct} $\bolds{K}$} &
\multicolumn{1}{c}{\textbf{error rate}} &
\multicolumn{1}{c@{}}{\textbf{reproducibility}} \\
\hline
{Separate K-means} & \phantom{00}{2} &{0.33 (0.001)} & {0.54 (0.07)}\\
{ } & \phantom{00}{0} & {0.33 (0.002)} & {0.47 (0.04)} \\
{Concatenated K-means} & { 100 } &
{ 0.01 (0.07) } & { 0.96 (0.03) } \\
{Separate sparse SVD} & \phantom{00}{0} &
{0.28 (0.10)} & {0.45 (0.03)}\\
{ } & \phantom{00}{0} & {0.31 (0.07)} & {0.44 (0.04)} \\
{Concatenated sparse SVD} & \phantom{0}{16} &
{0.01 (0.002)} & {0.59 (0.05)} \\
{Separate AHP-GMM} & \phantom{00}{0} &
{0.07 (0.13)} & {0.63 (0.05)}\\
{ } & \phantom{00}{0} &{0.32 (0.02)} & {0.54 (0.06)} \\
{Concatenated AHP-GMM} & {100} &
{0.01 (0.07)} & {0.98 (0.03)} \\
{Lasso iCluster} & {100} &
{0.0003 (0.001)} & {0.98 (0.01)} \\
{Enet iCluster} & {100} &
{0.0003 (0.001)} & {0.97 (0.02)} \\
{Fused lasso iCluster} & {100} &
{0 (0)} & {0.94 (0.05)} \\
\hline
\end{tabular*}
\end{table}
%

\begin{table}
\caption{Feature selection performance summarized over 50 simulated
data sets under $K=3$}\label{tab4}
\begin{tabular*}{\textwidth}{@{\extracolsep{\fill}}lcccc@{}}
\hline
& \multicolumn{2}{c}{\textbf{Data 1}} & \multicolumn{2}{c@{}}{\textbf{Data 2}}
\\[-4pt]
& \multicolumn{2}{c}{\hrulefill} & \multicolumn{2}{c@{}}{\hrulefill} \\
\multicolumn{1}{c}{} & \textbf{True} & \textbf{False} & \textbf{True}
& \textbf{False} \\
\multicolumn{1}{@{}l}{\textbf{Method}} & \textbf{positives} & \textbf{
positives} & \textbf{positives} & \textbf{positives} \\
\hline
{Separate K-means} & -- & -- & -- & --\\
{Concatenated K-means} & -- & -- & -- & --\\
{Separate sparse SVD} & 19.8 (0.7) & 349.6 (167.1) &
19.9 (0.3) & 347.5 (142.5) \\
{Concatenated sparse SVD} & 20 (0) & 396.6 (128.7) &
19.6 (1.6) & 395.4 (128.3) \\
{Separate AHP-GMM} & 15.8 (5.0) & 239.9 (245.5) &
15.5 (5.5) & 269.9 (246) \\
{Concatenated AHP-GMM} & 19.2 (1.7) & 0.33 (0.64) &
14.4 (4.0) & 0.21 (0.66) \\
{Lasso iCluster} & 20 (0) & 1.5 (1.4) & 19.9 (0.2) &
1.9 (1.5) \\
{Enet iCluster} & 20 (0) & 0.5 (0.6) & 19.8 (0.5) &
0.7 (1.0) \\
{Fused lasso iCluster} & 20 (0) & 0 (0) & 20 (0) & 0
(0) \\
\hline
\end{tabular*}
\end{table}

The core iCluster EM iterations are implemented in C. Table~\ref{tab5} shows
some typical computation times for problems of various dimensions on a
3.2~GHz Xeon Linux computer.

%
\begin{table}[b]
\caption{Computing time (in seconds) for typical runs of sparse
iCluster under various dimensions}\label{tab5}
\begin{tabular*}{\textwidth}{@{\extracolsep{\fill}}lcd{2.2}d{2.2}d{2.2}@{}}
\hline
& &\multicolumn{3}{c@{}}{\textbf{Time (in seconds)}}\\[-4pt]
& &\multicolumn{3}{c@{}}{\hrulefill}\\
$\bolds{p}$ & $\bolds{N}$ & \multicolumn{1}{c}{\textbf{Lasso iCluster}} & \multicolumn{1}{c}{\textbf{Elastic net iCluster}} &
\multicolumn{1}{c@{}}{\textbf{Fused lasso iCluster}} \\
\hline
\phantom{0}200 & 100 & 0.10 & 0.11 & 0.37\\
\phantom{0}500 & 100 & 0.50 & 0.36 & 3.56\\
1000 & 100 & 1.40 & 1.45 & 25.05\\
2000 & 100 & 6.49 & 5.90 & 76.40\\
5000 & 100 & 18.93 & 18.94 & \multicolumn{1}{c}{\phantom{000.}33 (min)}\\
\hline
\end{tabular*}
\end{table}

\section{Results}\label{sec7}
In this section we present details of two real data \mbox{applications}.
\subsection{Integration of epigenomic and transcriptomic profiling data
in the Holm breast cancer study}\label{sec7.1}
In Section~\ref{sec2} we discussed a motivating example using the \citet
{Pollack2002} data set. In this section we present our first real data
application which involves integrative analysis of DNA methylation and
gene expression data from the \citet{Holm} study. In this data set,
methylation profiling in 189 breast cancer samples using Illumina
methylation arrays for 1452 CpG sites (corresponding to 803
cancer-related genes) is available. The original study performed a
hierarchical clustering on the methylation data alone. Through manual
integration, the authors then correlated the methylation status with
gene expression levels for 511 oligonucleotide probes for genes with
CpG sites on the methylation assays in the same sample set. Here we
compare clustering of individual data types to various integration
approaches. We included the most variable 288 CpG sites (following a
similar procedure taken in the Holm study) in the methylation
data.\vadjust{\goodbreak}

%
\begin{figure}

\includegraphics{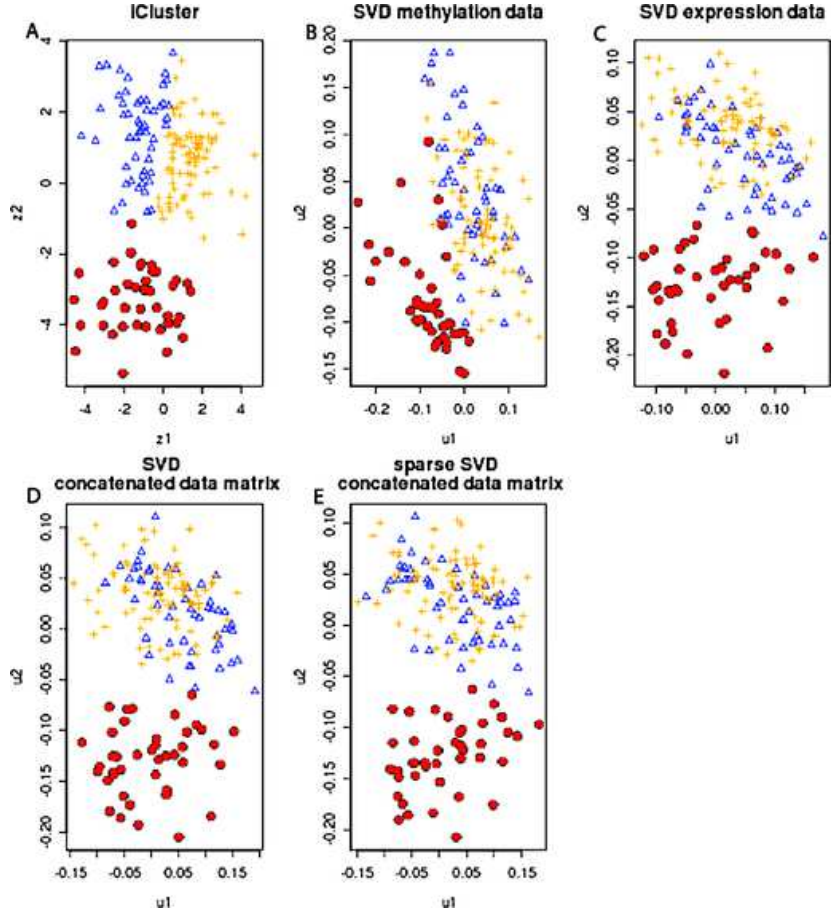}

\caption{Separation of the data points by \textup{(A)} latent variables from
sparse iCluster, \textup{(B)}~right singular vectors from SVD of the
methylation data alone, \textup{(C)} right singular vectors from SVD of the
expression data alone, \textup{(D)} SVD on the concatenated data matrix, and
\textup{(E)} sparse SVD on the concatenated data matrix. Red dots indicate
samples belonging to cluster 1, blue open triangles indicate samples
belonging to cluster 2, and orange pluses indicate samples belonging to
cluster 3.} \label{fig3}
\end{figure}

\begin{figure}

\includegraphics{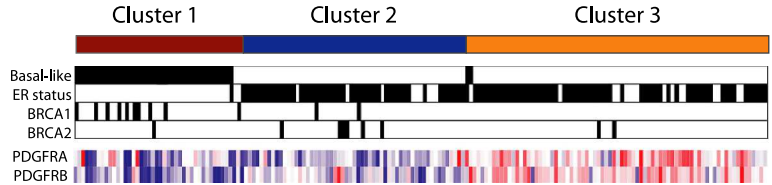}

\caption{Integrative clustering of the Holm study DNA methylation and
gene expression data revealed three clusters with a cross-validated
reproducibility of 0.7 and distinct clinical and molecular
characteristics.}\label{fig4}
\end{figure}

We applied sparse iCluster for a joint analysis of the methylation
($p_1=288$) and gene expression ($p_2=511$) data using different
penalty combinations. In Figure~\ref{fig3}(A) the first two latent variables
separated the samples into three distinct clusters. By associating the
cluster membership with clinical variables, it becomes clear that
tumors in cluster 1 are predominantly estrogen receptor (ER)-negative
and associated with the basal-like breast cancer subtype (Figure~\ref{fig4}).
Among the rest of the samples, sparse iCluster further identifies a
subclass (cluster 3) that highly expresses platelet-derived growth
factor receptors (\textit{PDGFRA/B}), which have been associated with
breast cancer progression [\citet{pdgfr}].\vadjust{\goodbreak}

%

%
\begin{table}[b]
\def\arraystretch{0.9}
\caption{Cluster reproducibility and number of genomic features
selected using\break sparse iCluster, sparse SVD on concatenated data matrix
and Adaptive Hierarchically Penalized Gaussian Mixture Model (AHP-GMM)
on concatenated data matrix. Two variations of the sparse iCluster
method were presented: iCluster (lasso, lasso) implements lasso penalty
for both data types, and iCluster (lasso, elastic net) implements lasso
penalty for the methylation data and elastic net penalty for the gene
expression data. K: the number of clusters. RI: reproducibility index}\label{tab6}
\begin{tabular*}{\textwidth}{@{\extracolsep{\fill}}lcccccc@{}}
\hline
&&\textbf{Selected}&\textbf{Selected}&&\textbf{Selected}&\textbf{Selected}\\
&&\textbf{methylation}&\textbf{expression}&&\textbf{methylation}&\textbf{expression}\\
$\bolds{K}$ & \textbf{RI} & \textbf{features}& \textbf{features}&\textbf{RI}
&\textbf{features} & \textbf{features} \\
\hline
& \multicolumn{3}{c}{iCluster (lasso, lasso)} & \multicolumn{3}{c@{}}{iCluster (lasso, elastic
net)}\\
2& 0.68 & 138 & 151 & 0.70 & 183 & 353 \\
3& 0.46 & 150 & 204 & 0.70 & 273 & 182 \\
4& 0.42 & 183 & 398 & 0.48 & 273 & 182 \\
5& 0.42 & 205 & 454 & 0.47 & 282 & 223 \\[6pt]
& \multicolumn{3}{c}{sparse SVD} & \multicolumn{3}{c}{AHP-GMM}\\
2& 0.78 & \phantom{00}1 & 105 & 0.93 & \phantom{00}9 & \phantom{0}63 \\
3& 0.34 & \phantom{00}1 & 134 & 0.42 & \phantom{0}28 & 105 \\
4& 0.27 & 288 & 511 & 0.49 & 116 & 368 \\
5& 0.22 & 273 & 504 & 0.43 & \phantom{0}42 & 243 \\
\hline
\end{tabular*}
\end{table}

In Section~\ref{sec3.2} we discussed an SVD approach on a combined data matrix
as a special case of our model. Here we present results from SVD and a
sparse SVD algorithm proposed by \citet{Witten2009a} on the
concatenated data matrix. Figures~\ref{fig3}(B) and~\ref{fig3}(C) indicate that SVD applied
to each data type alone can only separate one out of the three
clusters. Figures~\ref{fig3}(D) and~\ref{fig3}(E) indicate that data concatenation does not
perform any better in this analysis than separate analyses of each data
type alone.\looseness=-1

In Table~\ref{tab6} the results from sparse iCluster with two different sets of
penalty combinations are presented:\vadjust{\goodbreak} the combination of (lasso, lasso)
and the combination of (lasso, elastic net) for methylation and gene
expression data, respectively (Table~\ref{tab6} top panel). The reproducibility
index (RI) is computed for various $K$'s and penalty parameters are
sampled based on a uniform design described in Section~\ref{sec3.3}. As
described in Section~\ref{sec5}, RI (ranges between 0 and 1) measures the
agreement between the predicted cluster membership and the ``observed''
cluster membership using a 10-fold cross-validation.

%
\begin{figure}

\includegraphics{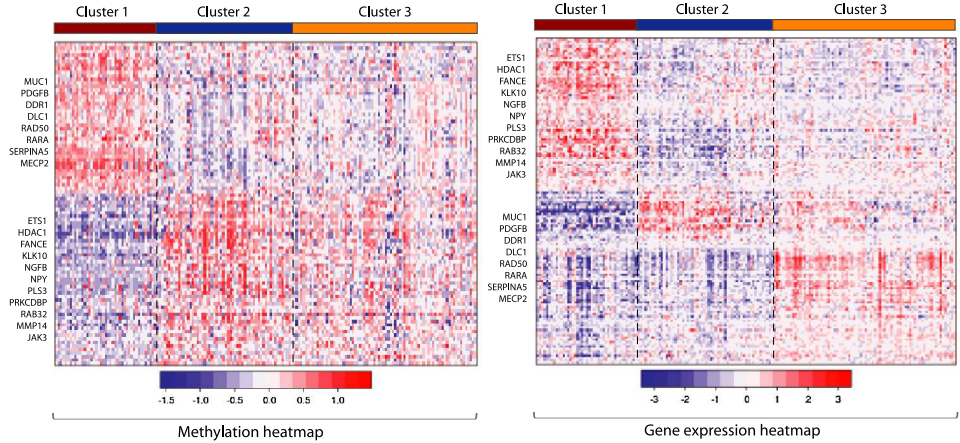}

\caption{Integrative clustering of the Holm study DNA methylation and
gene expression data revealed three clusters with a cross-validated
reproducibility of 0.7. Selected genes with negatively correlated
methylation and expression changes are indicated to the left of the
heatmap.}\label{fig5}
\end{figure}

Both methods identified a 2-cluster solution with an RI around 0.70,
distinguishing the ER-negative, Basal-like subtype from the rest of the
tumor samples (Figures~\ref{fig3} and~\ref{fig4}, samples labeled in
red). The iCluster (lasso, elastic net) method adds an $\ell_2$ penalty
term to encourage grouped selection of highly correlated genes in the
expression data. This approach further identified a 3-cluster solution
with high reproducibility ($\mathrm{RI}=0.70$). The additional division
finds a subgroup that highly expresses platelet-derived growth factor
receptors (Figure~\ref{fig4}).

Figure~\ref{fig5} displays heatmaps of the methylation and expression data.
Col\-umns are samples ordered by the integrated cluster assignment. Rows
are cluster-discriminating genes (with nonzero coefficient estimates)
grouped into gene clusters by hierarchical clustering. In total, there
are 273 differentially methylated genes and 182 differentially
expressed genes. Several cancer genes including \textit{MUC1\textup{,} SERPINA5\textup{,}
RARA\textup{,} MECP2 \textup{and} RAD50} are hypermethylated and show concordant
underexpression in cluster 1. On the other hand, hypomethylation of
cancer genes including \textit{ETS1\textup{,} HDAC1\textup{,} FANCE\textup{,} RAB32 \textup{and} JAK3} are
observed and, correspondingly, these genes show increased expression
levels.

To compare with other methods, we implemented the sparse SVD method by
\citet{Witten2009a} and an adaptive hierarchical penalized Gaussian
mixture model (AHP-GMM) by \citet{WangZhu2008} on the concatenated data
matrix. None of these methods generated additional insights beyond
separating the ER-negative and basal-like tumors from the others
(Figure~\ref{fig3} and Table~\ref{tab6}). Feature selection is predominantly ``biased''
toward gene expression features when directly applying sparse SVD on
the combined data matrix (bottom panel of Table~\ref{tab6}), likely due to the
larger between-cluster variances observed in the gene expression data.

%
\begin{figure}

\includegraphics{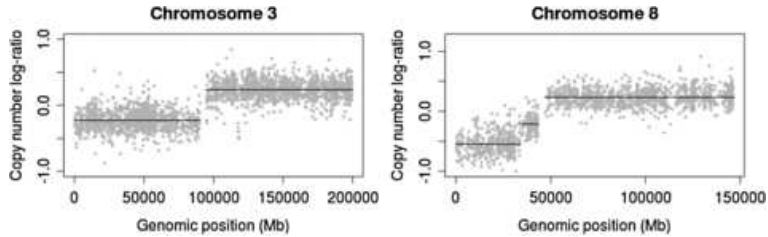}

\caption{Illustration of copy number probe-level data from a lung tumor
sample [Chitale et~al. (\citeyear{Ladanyi})]. Log-ratios of copy number (tumor versus
normal) on chromosomes 3 and 8 are displayed. Log-ratio greater than
zero indicates copy number gain and log-ratio below zero indicates
loss. Black line indicates the segmented value using the circular
binary segmentation method [Olshen et~al. (\citeyear{cbs1}), Venkatraman and Olshen (\citeyear{cbs2})].}\label{fig6}
\end{figure}

\subsection{Constructing a genome-wide portrait of concordant copy
number and gene expression pattern in a lung cancer data set}\label{sec7.2}
We applied the proposed method to integrate DNA copy number (aCGH data)
and mRNA expression data in a set of 193 lung adenocarcinoma samples
[\citet{Ladanyi}]. Figure~\ref{fig6} displays an example of the probe-level data
(log-ratios of tumor versus normal copy number) on chromosomes 3 and 8
in one tumor sample. Many samples in this data set display similar chr
3p whole-arm loss and chr 3q whole-arm gain.

Arm-length copy number aberrations are surprisingly common in cancer
[\citet{Beroukhim2010}], affecting up to thousands of genes within the
region of alteration. A broader challenge is thus to pinpoint the
``driver'' genes that have functional roles in tumor development from
those that are functionally neutral (``passengers''). To that end, an
integrative analysis with gene expression data could provide additional
insights. Genes that show concordant copy number and transcriptional
activities are more likely to have functional roles.

%
\begin{figure}

\includegraphics{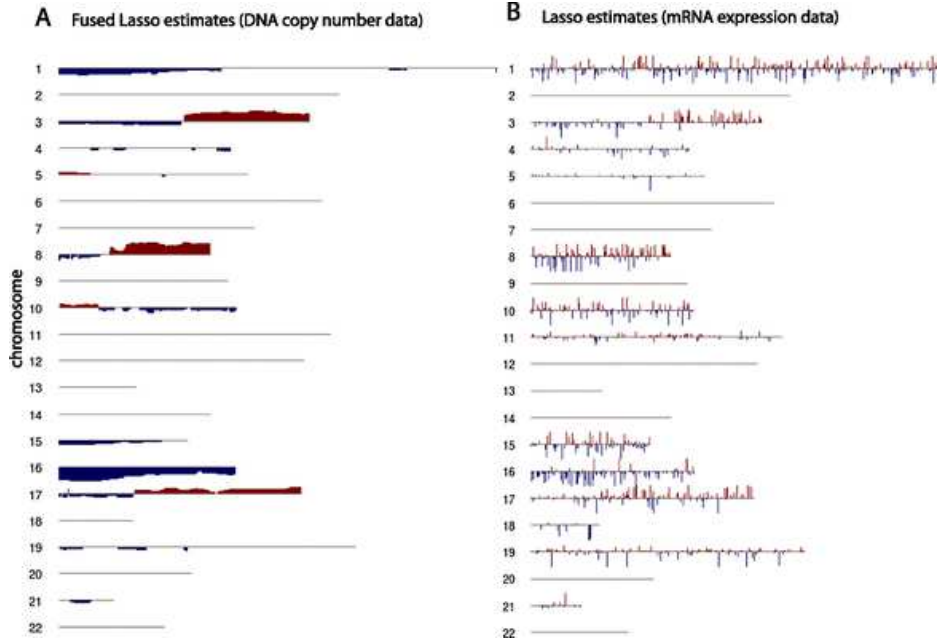}

\caption{Penalized coefficient vector estimates arranged by chromosomes
1 to 22 derived by iCluster (fused lasso, lasso) applied to the Chitale
et al. lung cancer data set. A single latent variable vector is used to
identify the major pattern of each chromosome.}\label{fig7}
\end{figure}

In the search for copy number-associated gene expression patterns, we
fit a sparse iCluster model for each of the 22 chromosomes using (fused
lasso, lasso) a~penalty combination for joint analysis of copy number
and gene expression data. To facilitate comparison, we compute a
2-cluster solution with a single latent variable vector $\VV z$ (instead
of estimating $K$) to extract the major pattern for each chromosome.
Penalty parameter tuning is performed as described before. In Figure~\ref{fig7}
we plot the 22 pairs of the sparse coefficient vectors ordered by
chromosomal position. The coefficients can be interpreted as the
difference between the two cluster means. Positive and negative
coefficient values in Figure~\ref{fig7}(A) thus indicate copy number gains and
losses in one cluster relative to the other. Similarly, in Figure~\ref{fig7}(B),
coefficient signs indicate over- or under-expression in one cluster
relative to the other. Concordant copy number and gene expression
changes can thus be directly visualized from Figure~\ref{fig7}.

Several chromosomes (1, 3, 8, 10, 15 and 16) show contiguous regions of
gains or losses spanning whole chromosome arms. As discussed before,
arm-length aberrations can affect up to thousands of genes within the
region of alteration. A great challenge is thus to pinpoint the
``driver'' genes that have important roles in tumor development from
those that are functionally neutral (``passengers''). To that end, an
integrative analysis could provide additional insights for identifying
potential drivers by revealing genes with concordant copy number and
transcriptional activities. Figure~\ref{fig7} shows that the application of the
proposed method can unveil a genome-wide pattern of such concordant
changes, providing a rapid way for identifying candidate genes of
biological significance. Several arm-level copy number alterations
(chromosomes 3, 8, 10, 16) exhibit concerted influence on the
expression of a small subset of the genes within the broad regions of
gains and losses.

\section{Discussion}\label{sec8}
Integrative genomics is a new area of research accelerated by
large-scale cancer genome efforts including the Cancer Genome Atlas
Project. New integrative analysis methods are emerging in this field.
\citet{Wessel1} proposed a nonparametric testing procedure for DNA copy
number induced differential mRNA gene expression. \citet{remMap2010}
and \citet{Paradigm2010} considered pathway and network analysis using
multiple genomic data sources. A number of others [\citet
{Waaijenborg2008, Parkhomenko2009, LeCao2009, Witten2009a, Witten2009b,
Soneson2010}] suggested using canonical correlation analysis (CCA) to
quantify the correlation between two data sets (e.g., gene expression
and copy number data). Most of this previous work focused on
integrating copy number and gene expression data, and none of these
methods were specifically designed for tumor subtype analysis.

We have formulated a penalized latent variable model for integrating
multiple genomic data sources. The latent variables can be interpreted
as a set of distinct underlying cancer driving factors that explain the
molecular phenotype manifested in the vast landscape of alterations in
the cancer genome, epigenome and transcriptome. Lasso, elastic net and
fused lasso penalty terms are used to induce sparsity in the feature
space. We derived an efficient and unified algorithm. The
implementation scales well for increasing data dimension.

A future extension on group-structured penalty terms is to incorporate
a grouping structure defined a priori. Two types of group
structures are relevant for our application. One is to treat the
$w_{i1},\ldots,w_{i(K-1)}$ as a group since they are associated with
the same feature. Yuan and Lin's group lasso penalty [\citet
{YuanLin2006}] can be applied directly. Similar to our current
algorithm, by using Fan and Li's local quadratic approximation, the
problem reduces to a ridge-type regression in each iteration. The other
extension is to incorporate the grouping structure among features to
boost the signal to noise ratio, for example, to treat the genes within
a pathway as a group. We can consider a hierarchical lasso penalty
[\citet{wang2009hierarchically}] to achieve sparsity at both the group
level and the individual variable level.

\section*{Acknowledgments}
We sincerely thank the Editor and the reviewers for the effort and care they
took in providing valuable comments and directions to improve the manuscript.
R. Shen and S. Wang contributed equally to this work.

%
%


\printaddresses

\end{document}